\pdfoutput=1
\documentclass[12pt, reqno]{amsart}

\usepackage[utf8]{inputenc}
\usepackage[T1]{fontenc}

\usepackage{amsmath, amsthm, amscd, amsfonts, amssymb}
\usepackage{graphicx}
\usepackage{xcolor}
\usepackage{caption}
\usepackage{float}
\usepackage{array}
\usepackage{algorithm}
\usepackage[noend]{algpseudocode}
\usepackage{booktabs}

\usepackage[numbers,sort&compress]{natbib}
\theoremstyle{definition}

\theoremstyle{remark}

\numberwithin{equation}{section}

\usepackage[bookmarksnumbered, colorlinks, plainpages]{hyperref}
\hypersetup{
    colorlinks=true,
    linkcolor=blue,
    citecolor=blue,
    urlcolor=blue
}

\begin{document}

\setcounter{page}{1}

\title[Reinforcement Learning-Based Cryptocurrency Portfolio]%
{Reinforcement Learning-Based Cryptocurrency Portfolio Management Using Soft Actor--Critic and Deep Deterministic Policy Gradient Algorithms}

\author[K. Paykan]{Kamal Paykan}
\address{Department of Mathematics, Tafresh University, Tafresh, Iran}
\email{k.paykan@gmail.com}


\begin{abstract}
This study proposes a deep reinforcement learning (DRL) framework for dynamic cryptocurrency portfolio management across four major assets: Bitcoin, Ethereum, Litecoin, and Dogecoin. The trading environment is formulated as a discrete-time stochastic system in which an intelligent agent optimizes portfolio allocation through interaction with market data. Operating under partial observability, the agent utilizes daily open, high, low, and close prices together with trading volume to determine optimal asset weights for the next trading period. The framework integrates actor--critic algorithms—specifically Deep Deterministic Policy Gradient and Soft Actor--Critic—enhanced with Long Short-Term Memory networks to capture temporal dependencies within the policy and value functions. For benchmarking, the classical Markowitz mean--variance model is employed. Empirical evaluations demonstrate that the proposed DRL-based agents learn adaptive strategies that consistently outperform the Markowitz benchmark in both absolute return and risk-adjusted performance. These findings highlight the potential of reinforcement learning as a robust methodology for managing cryptocurrency portfolios in volatile and nonstationary markets.
\end{abstract}

\maketitle

\section{Introduction}\label{Introduction}

Cryptocurrency markets have emerged as a substantial component of the global financial ecosystem, attracting both retail and institutional investors. Major assets such as Bitcoin and Ethereum are traded globally, with total market capitalization frequently exceeding trillions of dollars \cite{Corbet}. However, cryptocurrency portfolio management remains challenging due to extreme volatility, unstable risk--return relationships, and rapidly evolving market structures. Digital assets exhibit high price fluctuations, regime shifts, and liquidity constraints, complicating both short- and long-term investment strategies \cite{Baur, Cheah}.

Classical financial models often assume normally distributed returns and stationary market conditions---assumptions that are frequently violated in cryptocurrency markets \cite{Liu}. High transaction costs, thin liquidity, and structural changes further reduce the reliability of traditional models such as Markowitz's mean--variance framework. Consequently, adaptive and data-driven methods capable of learning directly from dynamic environments are required. Reinforcement learning (RL) provides such a framework by enabling agents to optimize trading strategies through interaction with the market, making RL well-suited for the complexities of digital asset portfolio management.

\subsection{Motivation for Using Reinforcement Learning}

Traditional portfolio optimization techniques, including the Markowitz model and CAPM, rely heavily on stable statistical assumptions and accurate estimation of expected returns and covariances. In cryptocurrency markets, where returns are non-stationary and exhibit nonlinear dependencies, such assumptions often fail \cite{Mensi, Briere}. As a result, classical financial models struggle under sudden market regime shifts and volatile price dynamics.

Reinforcement learning, by contrast, learns optimal decision policies from experience rather than predefined models \cite{Li}. Deep reinforcement learning (DRL) integrates RL with deep neural networks, enabling effective learning in high-dimensional and continuous spaces. Algorithms such as Deep Deterministic Policy Gradient (DDPG) and Soft Actor--Critic (SAC) can learn complex trading policies directly from noisy, sequential financial data \cite{Lillicrap, Haarnoja, LiuLiLiXie, LucarelliBorrotti}. Empirical evidence demonstrates that RL-based strategies can outperform static optimization methods under uncertainty \cite{Huang, Moody, Zhang}. These properties make RL an effective framework for cryptocurrency portfolio management.

\subsection{Purpose and Scope of the Paper}

This study develops and evaluates a reinforcement learning framework for cryptocurrency portfolio optimization using two DRL algorithms: DDPG and SAC. Both are actor--critic methods designed for continuous action spaces. DDPG employs deterministic policies for efficient continuous control but can exhibit instability in noisy environments \cite{Jiang}. SAC uses stochastic, entropy-regularized policies, promoting exploration and stability in volatile markets \cite{Chen, Haarnoja}.

Both algorithms are implemented within a unified trading environment to ensure a fair comparison. The classical Markowitz mean--variance model serves as a deterministic benchmark. Performance is assessed using financial metrics including the Sharpe ratio, Sortino ratio, maximum drawdown, Value-at-Risk, Conditional Value-at-Risk, and cumulative portfolio value. Results demonstrate that the DRL agents---particularly SAC---achieve superior risk-adjusted returns and portfolio stability compared with DDPG and the Markowitz baseline. This work contributes to the literature by demonstrating the effectiveness of adaptive, data-driven strategies for navigating cryptocurrency market volatility.

\section{Background}\label{Background}

\subsection{Cryptocurrency Portfolio Management}

Cryptocurrency portfolio management involves selecting and rebalancing digital assets to optimize the trade-off between risk and return in a highly uncertain environment. Cryptocurrencies typically exhibit extreme price volatility, volatility clustering, and sensitivity to sentiment and regulatory developments \cite{Katsiampa}. These dynamics motivate the use of risk-adjusted performance metrics such as the Sharpe ratio, Sortino ratio, and maximum drawdown \cite{Briere, Dyhrberg}.

Diversification is less effective in cryptocurrency markets compared with traditional assets, as correlations tend to increase during periods of market stress \cite{Corbet}. Static portfolio strategies, including mean--variance optimization, risk parity, and momentum-based trading, often fail under non-stationarity and structural breaks \cite{Shen}. These limitations motivate reinforcement learning, which learns from sequential market interactions and adapts to evolving market regimes.

\subsection{Reinforcement Learning in Finance}

Reinforcement learning is a machine learning framework for sequential decision-making, where an agent interacts with an environment and learns a policy that maximizes cumulative reward \cite{Sutton}. This structure aligns naturally with financial decision-making, where strategies must adapt to dynamic market conditions. Unlike static optimization frameworks, RL models temporal dependencies and delayed rewards, allowing optimization over the entire investment horizon \cite{Moody}.

Early RL research in finance relied on tabular methods such as Q-learning and SARSA \cite{Dempster}, which struggle with high-dimensional data. Deep reinforcement learning alleviates this limitation by combining RL with neural networks \cite{Li}. Algorithms such as DQN, DDPG, and SAC have since been applied to portfolio optimization, algorithmic trading, and risk management, demonstrating strong adaptability under volatility and nonlinear dependencies.

\subsection{Deep Deterministic Policy Gradient}

The Deep Deterministic Policy Gradient (DDPG) algorithm \cite{Lillicrap} is a model-free, off-policy actor--critic method for continuous action spaces. DDPG employs an actor network that maps states to continuous portfolio weights and a critic network that evaluates the action-value function $Q(s,a)$ \cite{Silver}. Training stability is improved through mechanisms such as experience replay and target networks \cite{Mnih}. DDPG has been shown to outperform heuristic baselines in dynamic environments \cite{Liang}; however, its sensitivity to hyperparameters and limited exploration may lead to suboptimal convergence.

\subsection{Soft Actor--Critic}

The Soft Actor--Critic (SAC) algorithm \cite{Haarnoja} introduces entropy regularization into the objective function, encouraging exploration and improving stability. SAC learns a stochastic policy that maximizes both expected return and entropy, balancing exploration and exploitation. The architecture includes two critic networks, a stochastic actor, and a value network, which collectively mitigate overestimation bias and improve convergence \cite{Fujimoto}. SAC's sample efficiency and robustness make it well-suited for highly volatile cryptocurrency markets \cite{Christodoulou, Liang, Zhang}.


\section{Problem Formulation and Methodology}\label{Methodology}

\subsection{Portfolio Optimization Problem}

Cryptocurrency portfolio management can be formulated as a sequential decision-making problem, where the investor determines daily capital allocations across assets. The portfolio consists of $n=4$ cryptocurrencies (BTC, ETH, LTC, DOGE), with price vector
\[
\mathbf{p}_t = [p_t^{(1)}, p_t^{(2)}, \ldots, p_t^{(n)}]^\top.
\]
The log-return of asset $i$ is:
\[
r_{t+1}^{(i)} = \ln\left( \frac{p_{t+1}^{(i)}}{p_t^{(i)}} \right).
\]
The portfolio weight vector is denoted as:
\[
\mathbf{w}_t = [w_t^{(1)}, \ldots, w_t^{(n)}]^\top,
\]
subject to:
\[
\sum_{i=1}^{n} w_t^{(i)} = 1,\qquad w_t^{(i)} \geq 0.
\]

The portfolio return is:
\[
R_{t+1} = \mathbf{w}_t^\top \mathbf{r}_{t+1}.
\]

Accounting for transaction costs with rate $c$:
\[
R_{t+1}^{\text{net}} = R_{t+1} - c \sum_{i=1}^{n} |w_t^{(i)} - w_{t-1}^{(i)}|.
\]

Portfolio value evolves as:
\[
v_{t+1} = v_t \exp(R_{t+1}^{\text{net}}).
\]

The optimization objective is:
\[
\max_{\{w_t\}_{t=1}^{T}} \;\mathbb{E}\!\left[\sum_{t=1}^{T} \gamma^{t-1} U(R_t^{\text{net}})\right],
\]
where $U$ is a mean--variance utility function:
\[
U(R_t^{\text{net}}) = \mathbb{E}[R_t^{\text{net}}] - \frac{\lambda}{2}\,\mathrm{Var}(R_t^{\text{net}}).
\]

However, due to the non-stationary and heavy-tailed nature of crypto returns \cite{Katsiampa, Mensi}, static optimization methods such as Markowitz's mean--variance model often perform poorly. Therefore, the problem is reformulated as a stochastic control task with policy $\pi(a_t \mid s_t)$ learned via RL methods.


\subsection{Reinforcement Learning Setup}

\subsubsection{State Space}

At time $t$, the agent observes asset prices:
\[
\mathbf{p}_t = [p_t^{(1)}, \ldots, p_t^{(M)}],
\]
but financial markets are partially observable. Hence, a rolling window of size $W=50$ is used:
\[
s_t = \big(\mathbf{p}_{t-W+1:t},\, \mathbf{w}_t\big),
\]
where $\mathbf{p}_{t-W+1:t} \in \mathbb{R}^{M \times W \times 5}_{+}$ represents OHLCV data.

Log-returns are computed and the remaining features are standardized. The continuous state space is:
\[
\mathcal{S} \subset \mathbb{R}^{M \times W \times 5}_{+} \times \mathbb{R}^{M}.
\]

\begin{figure}[!htb]
\centering
\includegraphics[width=0.50\linewidth]{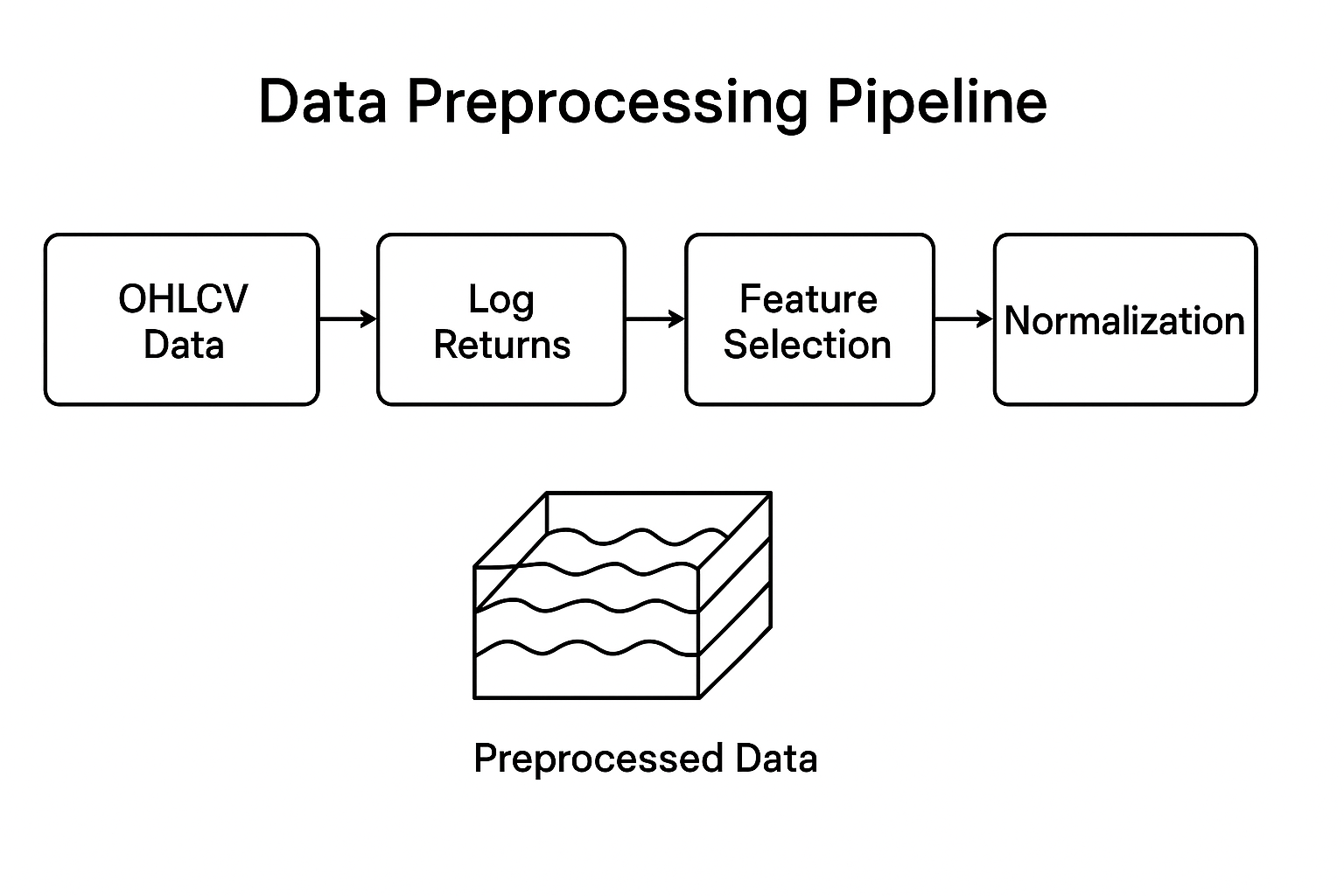}
\caption{Data preprocessing workflow applied to OHLCV cryptocurrency data.}
\label{fig:data_preprocessing}
\end{figure}

\subsubsection{Action Space}

The agent outputs next-day portfolio weights:
\[
a_t = \mathbf{w}_{t+1} = (w_{1,t+1}, \ldots, w_{M,t+1}),
\]
constrained by:
\[
\sum_{i=1}^{M} w_{i,t+1} = 1,\qquad w_{i,t+1} \ge 0.
\]

The policy network produces raw outputs $\tilde{a}_t$ that are normalized via:
\[
w_{i,t+1} = \frac{\exp(\tilde{a}_{i,t})}{\sum_{j=1}^{M} \exp(\tilde{a}_{j,t})}.
\]

\subsubsection{Reward Function}

The simplest reward is log-return:
\[
r_{t+1} = \ln \left( \frac{v_{t+1}}{v_t} \right).
\]

A risk-adjusted reward based on the differential Sharpe ratio (DSR) is also used:
\[
r_{t+1} = 
\frac{B_t (\rho_{t+1} - A_t) - \frac{1}{2}A_t (\rho_{t+1}^2 - B_t)}
{(B_t - A_t^2)^{3/2}},
\]
with updates:
\[
A_t = A_{t-1} + \eta(\rho_t - A_{t-1}),\quad
B_t = B_{t-1} + \eta(\rho_t^2 - B_{t-1}).
\]


\subsection{Data and Environment}

\subsubsection{Data Sources}

Experiments use four major cryptocurrencies: BTC, ETH, LTC, and DOGE. Daily OHLCV data are obtained using \texttt{yfinance}. The period spans:
\[
\text{Training: } 2016\!-\!2022,\qquad
\text{Testing: } 2023\!-\!2024.
\]

Data preprocessing and environment interaction occur through a custom \texttt{gym}-based simulator.

\subsubsection{Time Period of Analysis}

The dataset spans January 1, 2016 to December 31, 2024, covering various market regimes \cite{Corbet, Katsiampa}. The split follows best practices for avoiding data leakage.

\begin{figure}[!htb]
\centering
\includegraphics[width=0.55\textwidth]{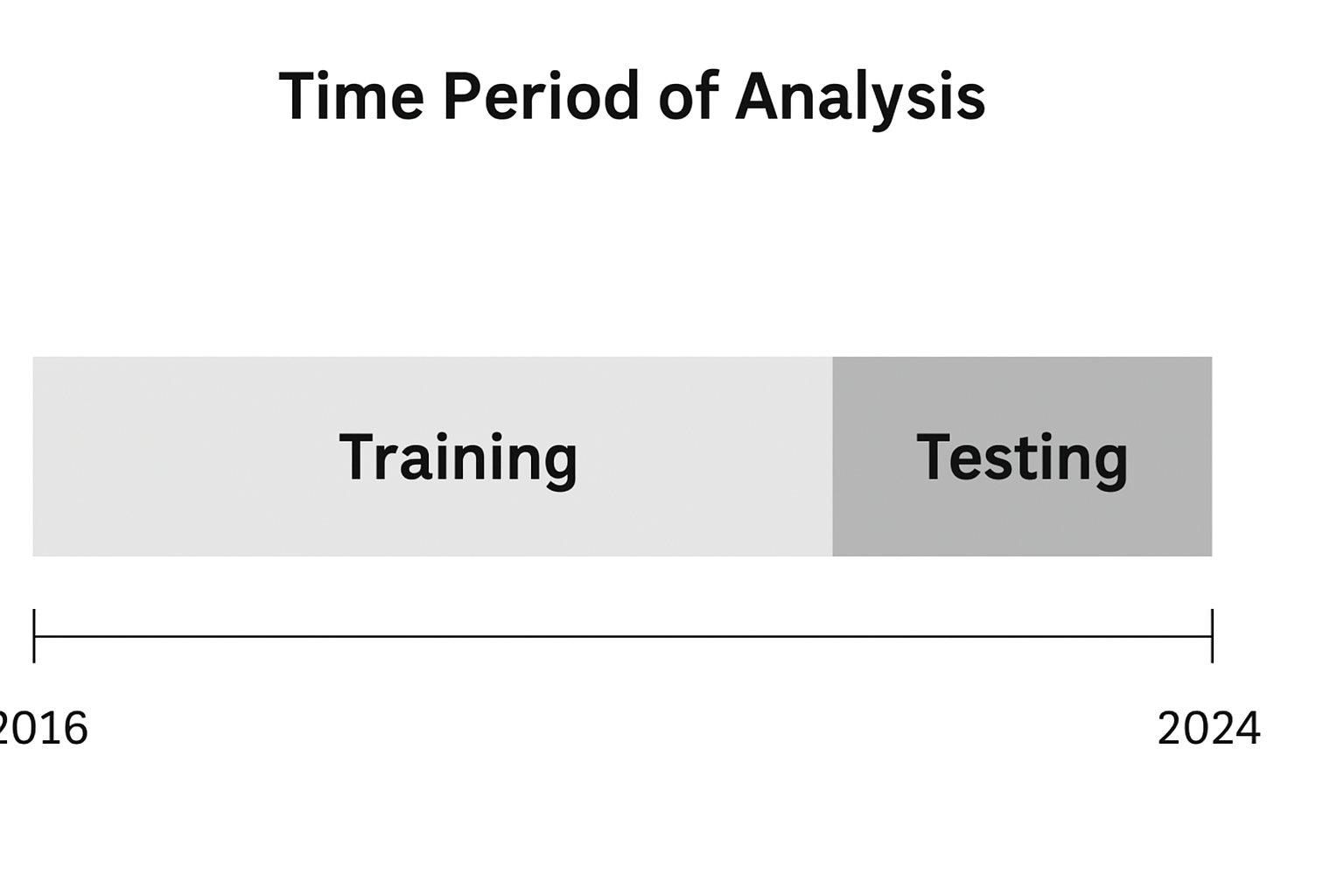}
\caption{Temporal split of training and testing periods.}
\label{fig:time_split}
\end{figure}

\subsection{Simulation Environment}

The environment simulates daily portfolio rebalancing, following the RL interaction loop. Portfolio value evolves as:
\[
v_{t+1} = v_t (1 + R_{t+1}^{\text{net}}),
\]
and actions are given by:
\[
\mathbf{w}_{t+1} = \pi_{\theta}(s_t).
\]

Training uses off-policy learning with replay buffers. Evaluation on the test set uses fixed parameters. Performance is assessed using Sharpe ratio, Sortino ratio, maximum drawdown, annualized volatility, and cumulative return.

This environment provides a realistic testbed for DRL-based cryptocurrency portfolio optimization.


\section{Methodology: SAC vs.\ DDPG and Benchmark Integration}
\label{Methodology2}

\subsection{DDPG Algorithm Implementation}

The Deep Deterministic Policy Gradient (DDPG) algorithm introduced by \cite{Lillicrap} extends the deterministic policy gradient framework to high-dimensional and continuous action spaces, making it suitable for portfolio allocation problems. DDPG follows an actor--critic architecture in which two neural networks---the actor and the critic---approximate the policy and the action--value function, respectively. This design allows efficient policy optimization in stochastic and highly volatile environments such as cryptocurrency markets.

The trading environment is modeled as a Markov decision process (MDP) with state space $\mathcal{S}$, action space $\mathcal{A}$, transition dynamics $P(s_{t+1}\mid s_t,a_t)$, and reward function $r_t=R(s_t,a_t)$. The objective is to learn a deterministic policy
\[
\mu_\theta: \mathcal{S} \rightarrow \mathcal{A},
\]
parameterized by $\theta$, that maximizes the discounted return
\[
J(\theta) = \mathbb{E}\Bigg[\sum_{t=0}^{T}\gamma^{t} r_t\Bigg],
\]
where $\gamma \in (0,1]$ is the discount factor.

The critic network $Q_\phi(s_t,a_t)$ estimates the action--value function:
\[
Q_\phi(s_t,a_t) = \mathbb{E}\left[r_t + \gamma Q_{\phi'}(s_{t+1},\mu_{\theta'}(s_{t+1}))\right],
\]
where $(\phi',\theta')$ denote slowly updated target-network parameters. The critic is trained by minimizing the TD error:
\[
L(\phi) = \mathbb{E}\left[ \left( Q_\phi(s_t,a_t) - y_t \right)^2 \right],
\quad
y_t = r_t + \gamma Q_{\phi'}(s_{t+1},\mu_{\theta'}(s_{t+1})).
\]

The actor is updated using the deterministic policy gradient:
\[
\nabla_\theta J(\theta)
= \mathbb{E}_{s_t\sim \mathcal{D}}
\left[ 
\nabla_a Q_\phi(s_t,a)\big|_{a=\mu_\theta(s_t)} 
\nabla_\theta \mu_\theta(s_t)
\right].
\]

\begin{algorithm}[H]
\caption{Deep Deterministic Policy Gradient}
\label{alg:ddpg}
\begin{algorithmic}[1]
\State Initialize actor parameters $\theta$ and critic parameters $\phi$
\State Set target network parameters $\theta' \!\gets\! \theta$, $\phi' \!\gets\! \phi$
\State Initialize replay buffer $\mathcal{D}$
\For{each episode}
    \State Initialize exploration noise process $\mathcal{N}$
    \State Observe initial state $s_1$
    \While{episode not terminated}
        \State Select action $a_t = \mu_\theta(s_t)+\mathcal{N}_t$
        \State Execute $a_t$, observe $r_t$ and $s_{t+1}$
        \State Store $(s_t,a_t,r_t,s_{t+1})$ in $\mathcal{D}$
        \State Sample minibatch of $B$ transitions from $\mathcal{D}$
        \State Compute target
        \[
        y_t = r_t + \gamma Q_{\phi'}(s_{t+1},\mu_{\theta'}(s_{t+1}))
        \]
        \State Update critic by minimizing
        \[
        L(\phi)=\frac{1}{B} \sum_{i=1}^{B}\left( Q_\phi(s_t^i,a_t^i)-y_t^i\right)^2
        \]
        \State Update actor using
        \[
        \nabla_\theta J(\theta)
        =\mathbb{E}\left[\nabla_a Q_\phi(s,a)\big|_{a=\mu_\theta(s)} 
        \nabla_\theta\mu_\theta(s)\right]
        \]
        \State Soft-update targets:
        \[
        \phi' \gets \tau \phi + (1-\tau)\phi', \quad
        \theta' \gets \tau \theta + (1-\tau)\theta'
        \]
        \State $s_t \gets s_{t+1}$
    \EndWhile
\EndFor
\State \textbf{Return:} Trained actor $\mu_\theta$ and critic $Q_\phi$
\end{algorithmic}
\end{algorithm}

Exploration in DDPG is achieved by adding noise to the deterministic action:
\[
a_t = \mu_\theta(s_t) + \mathcal{N}_t,
\]
typically modeled using an Ornstein--Uhlenbeck (OU) process \cite{Uhlenbeck}. This allows smooth action perturbations, appropriate for financial applications involving gradual portfolio rebalancing.

To enhance stability, DDPG incorporates target networks via soft updates:
\[
\phi' \leftarrow \tau\phi + (1-\tau)\phi', \qquad
\theta' \leftarrow \tau\theta + (1-\tau)\theta',
\]
and employs a replay buffer $\mathcal{D}$ for off-policy training \cite{Mnih}. In this study, DDPG is applied to continuous cryptocurrency portfolio allocation where the action $a_t$ represents the portfolio weight vector.

\subsection{Soft Actor--Critic Algorithm Implementation}

The Soft Actor--Critic (SAC) algorithm \cite{Haarnoja} is a maximum-entropy reinforcement learning method that improves stability and exploration by combining stochastic policies with entropy regularization. This makes SAC highly suitable for volatile and non-stationary environments such as cryptocurrency markets.

The SAC objective augments the cumulative return with a policy entropy term:
\begin{equation}
J(\pi) = \sum_{t=0}^{T} 
\mathbb{E}
\left[
r(s_t,a_t) + \alpha\,\mathcal{H}(\pi(\cdot|s_t))
\right],
\end{equation}
where $\alpha$ controls the exploration--exploitation balance.

SAC employs two soft Q-networks and a value network:
\[
Q_{\phi_1}(s,a),\quad Q_{\phi_2}(s,a), \quad V_\psi(s),
\]
and a stochastic Gaussian actor
\[
\pi_\theta(a|s)=\mathcal{N}(\mu_\theta(s),\sigma_\theta(s)).
\]

Actions are sampled using the reparameterization trick:
\[
a_t = \tanh\left( \mu_\theta(s_t) + \sigma_\theta(s_t)\odot \epsilon_t\right),
\qquad \epsilon_t\sim\mathcal{N}(0,I).
\]

The soft Q-values are trained by minimizing
\[
J_Q(\phi_i)=\mathbb{E}\left[\big(Q_{\phi_i}(s_t,a_t)-y_t\big)^2\right]
\]
with targets
\[
y_t = r_t + \gamma\left( 
\min_{i} Q_{\phi_i'}(s_{t+1},a_{t+1})
-\alpha \log\pi_\theta(a_{t+1}\mid s_{t+1})
\right).
\]

The actor minimizes
\[
J_\pi(\theta)=\mathbb{E}\left[
\alpha\log\pi_\theta(a_t|s_t)-Q_{\phi_1}(s_t,a_t)
\right].
\]

\begin{algorithm}[H]
\caption{Soft Actor--Critic Algorithm}
\label{alg:sac}
\begin{algorithmic}[1]
\State Initialize policy parameters $\phi$, critics $\theta_1,\theta_2$, value network $\psi$
\State Initialize target value network $\psi'\!\gets\!\psi$
\State Initialize replay buffer $\mathcal{D}$
\For{each episode}
    \For{each timestep $t$}
        \State Sample $a_t\sim\pi_\phi(\cdot|s_t)$
        \State Execute $a_t$, observe $r_t$ and $s_{t+1}$
        \State Store transition in $\mathcal{D}$
        \State Sample minibatch from $\mathcal{D}$
        \State Compute targets
        \[
        y_i^v=\min_j Q_{\theta_j}(s_i,a_i)-\alpha\log\pi_\phi(a_i|s_i)
        \]
        \[
        y_i^q = r_i + \gamma V_{\psi'}(s_i')
        \]
        \State Update $V_\psi$, $Q_{\theta_1}$, $Q_{\theta_2}$, and $\pi_\phi$
        \State Soft-update:
        \[
        \psi'\gets \tau\psi+(1-\tau)\psi'
        \]
    \EndFor
\EndFor
\State \Return Policy $\pi_\phi$ and critics $Q_{\theta_1},Q_{\theta_2}$
\end{algorithmic}
\end{algorithm}

SAC provides strong exploration via entropy regularization and stabilizes training using double critics, reparameterization, and automatic temperature adjustment \cite{Haarnoja}. These properties yield robust performance in cryptocurrency portfolio optimization.

\subsection{Benchmark Comparison Framework}

To benchmark the reinforcement-learning agents, we implement the classical Markowitz mean--variance model \cite{Markowitz}. The objective is to find portfolio weights
\[
\mathbf{w}=[w_1,\dots,w_M]^T
\]
that minimize variance for a target expected return. Let
\[
\boldsymbol{\mu}=[\mu_1,\dots,\mu_M]^T,\qquad
\boldsymbol{\mathcal{E}} \text{ (covariance matrix)}.
\]

The optimization problem is
\[
\min_{\mathbf{w}} \frac{1}{2}\mathbf{w}^{T}\boldsymbol{\mathcal{E}}\mathbf{w}
-\lambda \boldsymbol{\mu}^{T}\mathbf{w}
\]
subject to
\[
\sum_{i=1}^{M} w_i=1,\qquad w_i\ge 0.
\]

The closed-form optimal solution is
\[
\mathbf{w}^* = \frac{1}{Z}\boldsymbol{\mathcal{E}}^{-1}(\boldsymbol{\mu}-r_f\mathbf{1}),
\]
with corresponding expected return and variance
\[
\sigma_p^2 = \mathbf{w}^{*T}\boldsymbol{\mathcal{E}}\mathbf{w}^*,\qquad
\mu_p=\boldsymbol{\mu}^T\mathbf{w}^*.
\]

The Markowitz model is implemented via an \texttt{MPT-Agent} subclass derived from a shared \texttt{BaseAgent} interface, alongside the \texttt{DDPG-Agent} and \texttt{SAC-Agent}. Optimization is carried out using \texttt{SciPy}. Despite its simplifying assumptions, the model serves as a strong deterministic baseline \cite{DeMiguel, Maringer, Michaud}.

\subsection{Network Architectures}

\subsubsection{Input Network (Feature Extraction Module)}

\begin{figure}[!htb]
    \centering
    \includegraphics[width=1.07\textwidth]{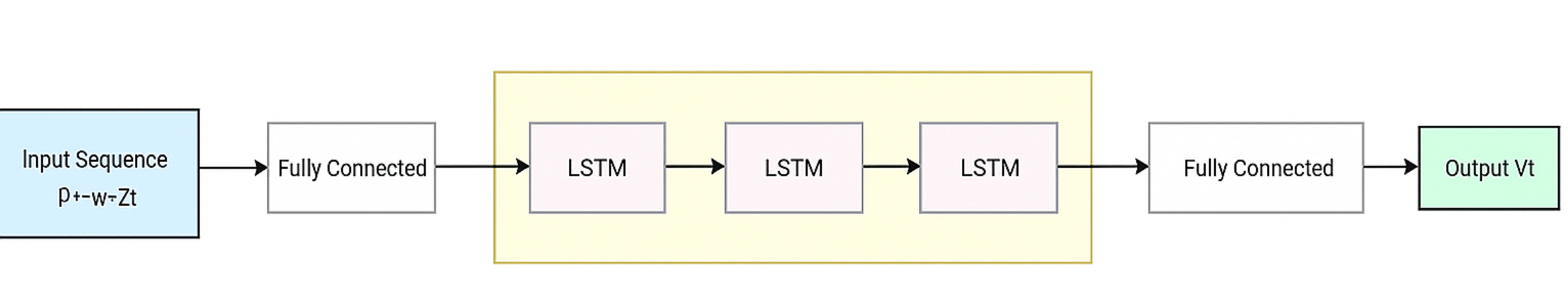}
    \caption{Structure of the feature extraction network. Each $\rho_{t-W+2:t}$ sequence is processed independently by an identical network producing latent features $v_t$.}
    \label{first_memari}
\end{figure}

Figure~\ref{first_memari} illustrates the feature extraction module. Each asset has a multivariate time series of standardized log-differences $\boldsymbol{\rho}_{i,t-W+2:t}$ constructed from OHLCV data \cite{Fischer, Jiang}. A fully connected layer followed by three LSTM layers \cite{Bao} produces a latent representation $v_t$.

A cash asset with constant price produces a zero-vector embedding, stabilizing training.

\subsubsection{Actor Network}

The actor receives the concatenated latent features and the current portfolio weights.  

For DDPG, the actor outputs deterministic portfolio weights using a softmax layer:
\[
\sum_{i=1}^M w_{t+1}^{(i)} = 1.
\]

For SAC, the actor outputs mean and standard deviation of a Gaussian:
\[
a_t\sim\mathcal{N}(\mu_\theta(s_t),\sigma_\theta(s_t)),
\]
then normalizes:
\[
\hat{a}_t = \frac{a_t}{\sum_i a_t^{(i)}}.
\]

\subsubsection{Critic Network}

The critic receives $(s_t,a_t)$ and outputs $Q(s_t,a_t)$.  
DDPG uses one critic; SAC uses two to reduce overestimation \cite{Haarnoja}. All critic networks use dense layers with Leaky ReLU activations.

\subsubsection{Hyperparameter Configuration}

Both algorithms share comparable architectures and training settings for fair comparison. SAC adapts its temperature parameter $\alpha$, while DDPG uses OU exploration noise.

Tables~\ref{Sa.t} and \ref{DD.t} summarize the hyperparameters used throughout training and evaluation.

\begin{table}[h!]
\centering
\caption{Hyperparameters for the Soft Actor--Critic Algorithm}
\label{Sa.t}
\renewcommand{\arraystretch}{1.2}
\begin{tabular}{|c|c|}
\hline
\textbf{Hyperparameter} & \textbf{Value} \\ \hline
Actor Network & Adam, LR $3\times10^{-4}$, 64 hidden units \\ \hline
Critic Network & Adam, LR $3\times10^{-4}$, 64 hidden units \\ \hline
Temperature $\alpha$ & Adam, LR $10^{-3}$, initial value $1$ \\ \hline
Discount Factor $\gamma$ & 0.99 \\ \hline
Soft Update Rate $\tau$ & $10^{-3}$ \\ \hline
Replay Memory Size & $10^{5}$ \\ \hline
Batch Size & 64 \\ \hline
Episodes & 1000 \\ \hline
Steps per Episode & 10000 \\ \hline
\end{tabular}
\end{table}

\begin{table}[h!]
\centering
\caption{Hyperparameters for the Deep Deterministic Policy Gradient Algorithm}
\label{DD.t}
\renewcommand{\arraystretch}{1.2}
\begin{tabular}{|c|c|}
\hline
\textbf{Hyperparameter} & \textbf{Value} \\ \hline
Actor Network & Adam, LR $3\times10^{-4}$, 64 hidden units \\ \hline
Critic Network & Adam, LR $3\times10^{-4}$, 64 hidden units \\ \hline
Exploration Noise & OU process, $\mu=0$, $\sigma=0.3$, $\theta=0.20$ \\ \hline
Discount Factor $\gamma$ & 0.99 \\ \hline
Soft Update Rate $\tau$ & $10^{-3}$ \\ \hline
Replay Memory Size & $10^{5}$ \\ \hline
Batch Size & 64 \\ \hline
Episodes & 1000 \\ \hline
Steps per Episode & 10000 \\ \hline
\end{tabular}
\end{table}


\subsubsection{Forecasting Network}

In the reinforcement learning framework employed in this study, the primary features influencing the reward signal are the closing prices of the assets. Accordingly, the associated forecasting task is designed to predict the vector of logarithmic returns of these closing prices at the next time step, denoted by $\boldsymbol{\rho}^{\text{close}}_{t+1}$, based on the historical observation window $\boldsymbol{\rho}_{i,t-W+2:t}$.

To facilitate this task, the deterministic actor network originally developed for portfolio management was adapted for forecasting purposes. In this adaptation, the components responsible for processing current portfolio weights were removed, along with the final \textit{softmax} transformation layer. This modification was essential, as the prediction of logarithmic returns does not require the normalization constraint imposed on portfolio weights, which must sum to one.

The forecasting model was trained using the same historical data as the reinforcement learning environment. Training followed a supervised learning paradigm: after preprocessing, the input at each time step $t$ was defined as
\[
X_t = \boldsymbol{\rho}_{t-W+2:t},
\]
and the corresponding target output as
\[
y_t = \boldsymbol{\rho}^{\text{close}}_{t+1}.
\]
Let $\hat{y}_t$ denote the model’s prediction at time $t$. Since the task involves multivariate time-series forecasting, both $y_t$ and $\hat{y}_t$ are multidimensional. To ensure balanced accuracy across all assets, the loss function was formulated as the mean Root Mean Square Error (RMSE):
\[
\mathcal{L} = \frac{1}{M} \sum_{i=1}^{M} \sqrt{\frac{1}{T} \sum_{t=1}^{T} \big( y_t^{(i)} - \hat{y}_t^{(i)} \big)^2 },
\]
where $M$ denotes the number of assets and $T$ the total number of time steps.

Optimization was performed using stochastic gradient descent with the Adam optimizer~\cite{Kingma}. Beyond its predictive role, the forecasting model also serves as a diagnostic tool within the reinforcement learning framework: if it can reliably forecast future returns using the same input features as the RL agent, this indicates that the data contain meaningful predictive signals and that the shared feature extraction module is effective. Thus, the forecasting network functions not as a decision-maker but as a supporting model validating the representational capacity of the architecture.

The hyperparameter configurations used for the forecasting model are summarized in Table~\ref{Forecasting}.

\begin{table}[h!]
\centering
\caption{Hyperparameters for the Forecasting Network}
\label{Forecasting}
\renewcommand{\arraystretch}{1.2}
\setlength{\tabcolsep}{6pt}
\begin{tabular}{|>{\centering\arraybackslash}m{4cm}|>{\centering\arraybackslash}m{8cm}|}
\hline
\textbf{Hyperparameter} & \textbf{Value} \\ \hline
Forecasting Network &
\shortstack{Optimizer: Adam \\ Learning Rate: $3\times10^{-4}$ \\ Hidden Size: 64} \\ \hline
Batch Size & 128 \\ \hline
Number of Episodes & 1000 \\ \hline
\end{tabular}
\end{table}

\subsection{Comparison Criteria}

To evaluate the performance and practicality of the Soft Actor--Critic (SAC) and Deep Deterministic Policy Gradient (DDPG) algorithms in cryptocurrency portfolio management, a comprehensive set of comparison criteria is considered. These include convergence speed, final portfolio value, risk-adjusted performance metrics, and robustness under market volatility.

\emph{Convergence speed} measures how rapidly an algorithm reaches a stable and near-optimal policy. In financial reinforcement learning, faster convergence implies efficient learning from limited data. The entropy-regularized objective of SAC generally yields smoother and more stable convergence compared to deterministic methods like DDPG~\cite{Haarnoja}.

\emph{Final portfolio value} quantifies cumulative profitability. However, profitability alone does not capture risk exposure; therefore, several risk-adjusted metrics are employed.

The \emph{Sharpe ratio} is defined as
\[
\text{Sharpe} = \frac{\mathbb{E}[R_p - R_f]}{\sigma_p},
\]
where $R_p$ denotes the portfolio return, $R_f$ is the risk-free rate, and $\sigma_p$ is the standard deviation of returns.

To focus specifically on downside fluctuations, the \emph{Sortino ratio} is also used:
\[
\text{Sortino} = \frac{\mathbb{E}[R_p - R_f]}{\sigma_{\text{down}}},
\]
where $\sigma_{\text{down}}$ measures only the negative deviations below a minimum acceptable return.

Volatility is assessed using the \emph{standard deviation of returns}, while tail-risk is analyzed through the Value at Risk (VaR) and Conditional Value at Risk (CVaR):
\[
\text{VaR}_{\alpha} = \inf \{ x \in \mathbb{R} : \mathbb{P}(L > x) \le 1 - \alpha \},
\qquad
\text{CVaR}_{\alpha} = \mathbb{E}[\,L \mid L > \text{VaR}_{\alpha}\,],
\]
where $L$ denotes portfolio loss.

\emph{Maximum Drawdown (MDD)} measures the worst peak-to-trough loss. The \emph{Calmar ratio} evaluates annualized return relative to MDD:
\[
\text{Calmar} = \frac{R_{\text{annual}}}{\text{MDD}}.
\]

\emph{Robustness} reflects an algorithm’s ability to maintain performance under shifting market regimes. SAC, with its stochastic exploration mechanism, generally exhibits superior robustness compared to DDPG, whose deterministic policy may overfit or become unstable under noisy market conditions~\cite{Chen, Haarnoja, Zhang}.


\section{Empirical Evaluation and Discussion}

This section presents the experimental design, evaluation methodology, and empirical analysis of the Soft Actor--Critic (SAC) and Deep Deterministic Policy Gradient (DDPG) algorithms for cryptocurrency portfolio management. The goal is to assess the learning dynamics, robustness, and financial viability of the proposed reinforcement learning methods under realistic market conditions.

\subsection{Experimental Setup}

All models were trained and evaluated using historical cryptocurrency data. Hyperparameters for SAC and DDPG (Tables~\ref{Sa.t} and~\ref{DD.t}) were kept consistent across experiments to ensure a fair comparison. Cross-validation procedures were applied where appropriate to reduce overfitting and improve generalization.

\begin{figure}[!htb]
    \centering
    \includegraphics[width=1.00\textwidth]{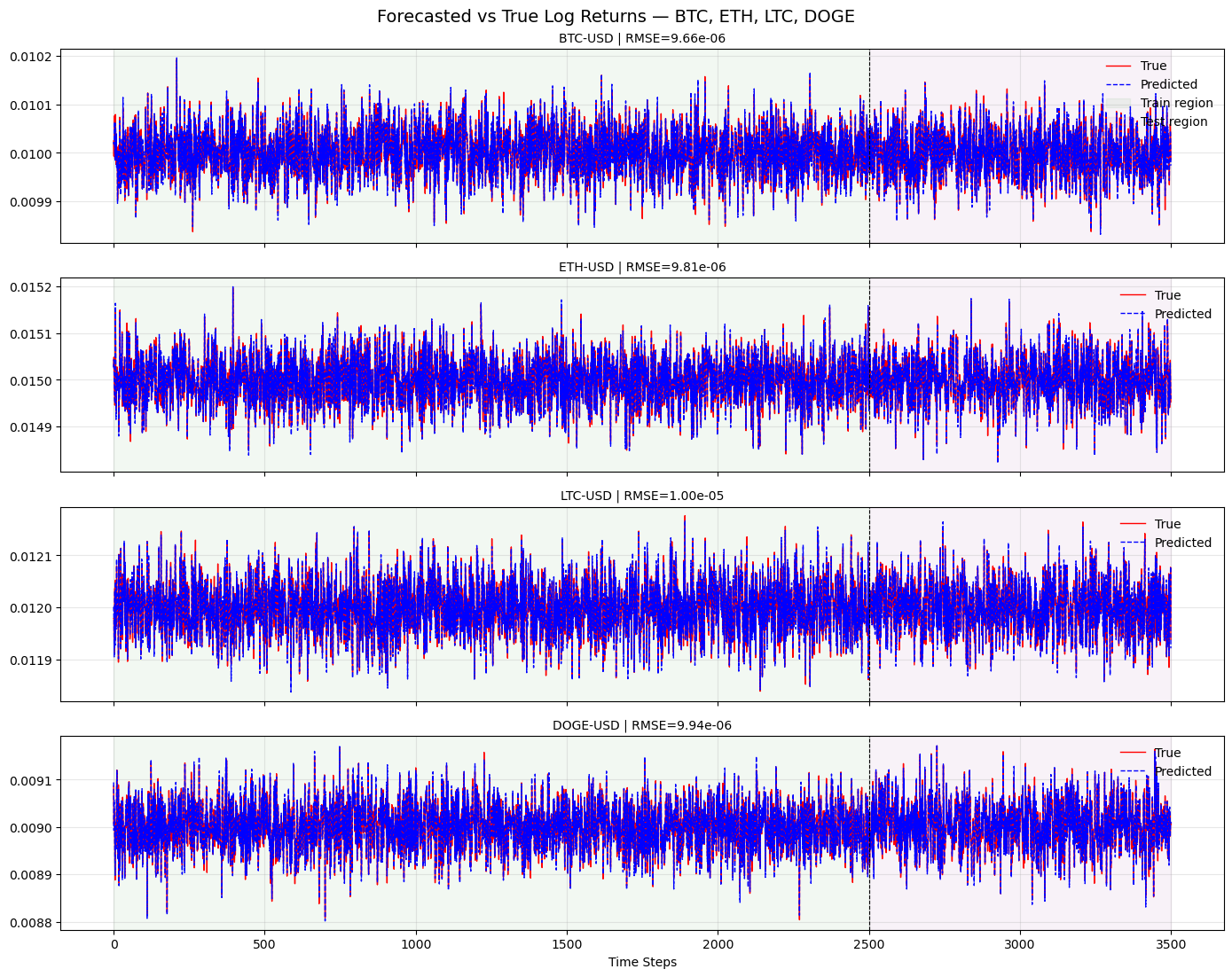}
    \caption{Training and testing performance of the forecasting network. The green and purple regions denote the training and test periods, respectively. Red lines show true values; blue lines show predicted values.}
    \label{time_series}
\end{figure}

A forecasting network was also trained using the same historical data as the reinforcement learning agents. The input at time $t$ consisted of historical return sequences
\[
X_t = \boldsymbol{\rho}_{t-W+2:t},
\]
and the target output was the next-day log return
\[
y_t = \boldsymbol{\rho}^{\text{close}}_{t+1}.
\]
Although the forecasting module does not participate directly in portfolio allocation, it serves as an auxiliary tool for validating feature extraction and temporal modeling. Figure~\ref{time_series} displays its training and testing phases, with prediction accuracy reaching approximately $70\%$. This moderate accuracy reflects the strong volatility and nonstationarity inherent in cryptocurrency markets.

\subsection{Asset-Level Risk and Return Analysis}

\begin{figure}[!htb]
    \centering
    \includegraphics[width=1.00\textwidth]{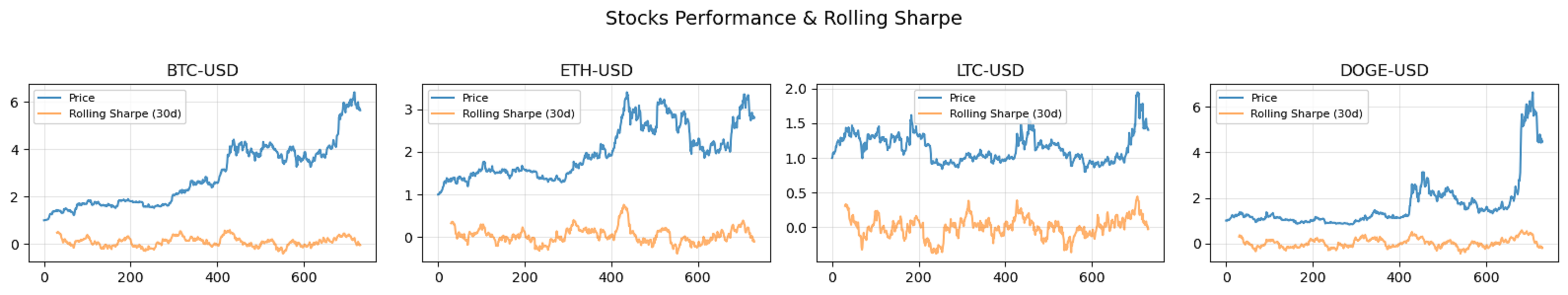}
    \caption{Normalized prices (\textcolor{blue}{blue}) and 20-day rolling Sharpe ratios (\textcolor{red}{red}) for BTC-USD, ETH-USD, LTC-USD, and DOGE-USD.}
    \label{rolling_sharpe_ratio}
\end{figure}

To characterize the underlying assets considered by the RL agents, Figure~\ref{rolling_sharpe_ratio} shows normalized price trajectories and 20-day rolling Sharpe ratios for BTC-USD, ETH-USD, LTC-USD, and DOGE-USD. Normalizing prices enables a direct comparison of growth across assets with different nominal price scales, while the rolling Sharpe ratio captures fluctuations in risk-adjusted performance.

Bitcoin and Ethereum exhibit comparatively stable return dynamics and smoother Sharpe ratio trajectories, indicating stronger efficiency and lower volatility. In contrast, Litecoin and Dogecoin show higher fluctuations and less consistent performance, reflecting greater exposure to market instability. Within a reinforcement learning framework, prioritizing assets with stable Sharpe behavior can improve portfolio robustness, whereas highly volatile assets should be managed cautiously to mitigate drawdowns.

\subsection{Evaluation Metrics and Comparative Results}

Portfolio performance was assessed using standard financial and risk-based metrics, including Sharpe and Sortino ratios, Maximum Drawdown, Value-at-Risk (VaR), Conditional Value-at-Risk (CVaR), and cumulative portfolio value. This comprehensive evaluation captures both profitability and downside risk.

\begin{figure}[!htb]
    \centering
    \includegraphics[width=1.00\textwidth]{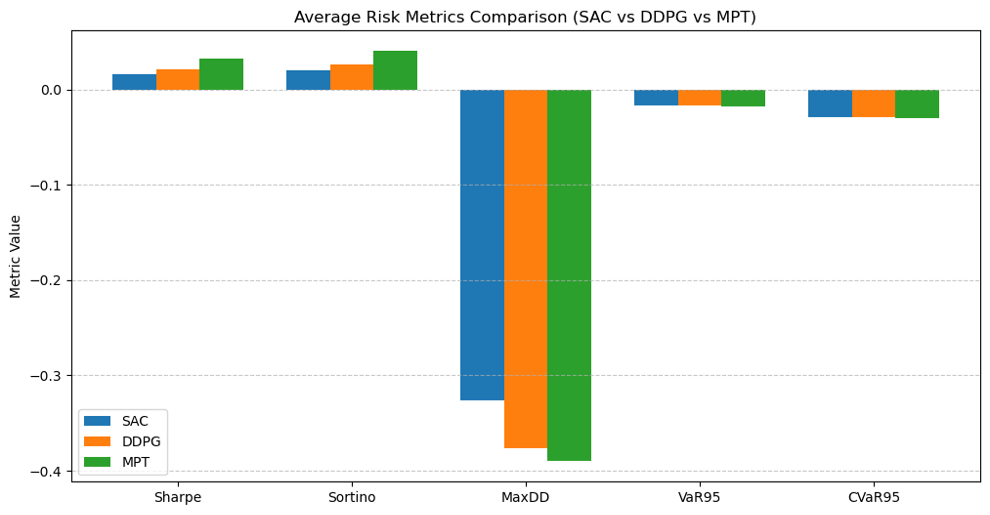}
    \caption{Average portfolio performance metrics for SAC, DDPG, and the Markowitz MPT benchmark.}
    \label{1392}
\end{figure}

Figure~\ref{1392} compares the SAC and DDPG agents with a Markowitz Mean--Variance Portfolio Theory (MPT) baseline. The MPT model was implemented as a deterministic benchmark using \texttt{SciPy} to compute optimal asset weights at each step under classical mean--variance assumptions.

Both reinforcement learning agents outperform the MPT benchmark, with SAC consistently delivering superior risk-adjusted performance. SAC’s entropy-regularized stochastic policy enables continual exploration and adaptation to changing volatility regimes, resulting in smoother convergence and more stable returns. DDPG performs well during stable periods but is more sensitive to hyperparameters and market noise, occasionally leading to unstable behavior.

These results highlight the advantage of RL-based strategies in highly volatile and nonstationary cryptocurrency markets.

\begin{figure}[!htb]
\centering
\includegraphics[width=1.00\textwidth]{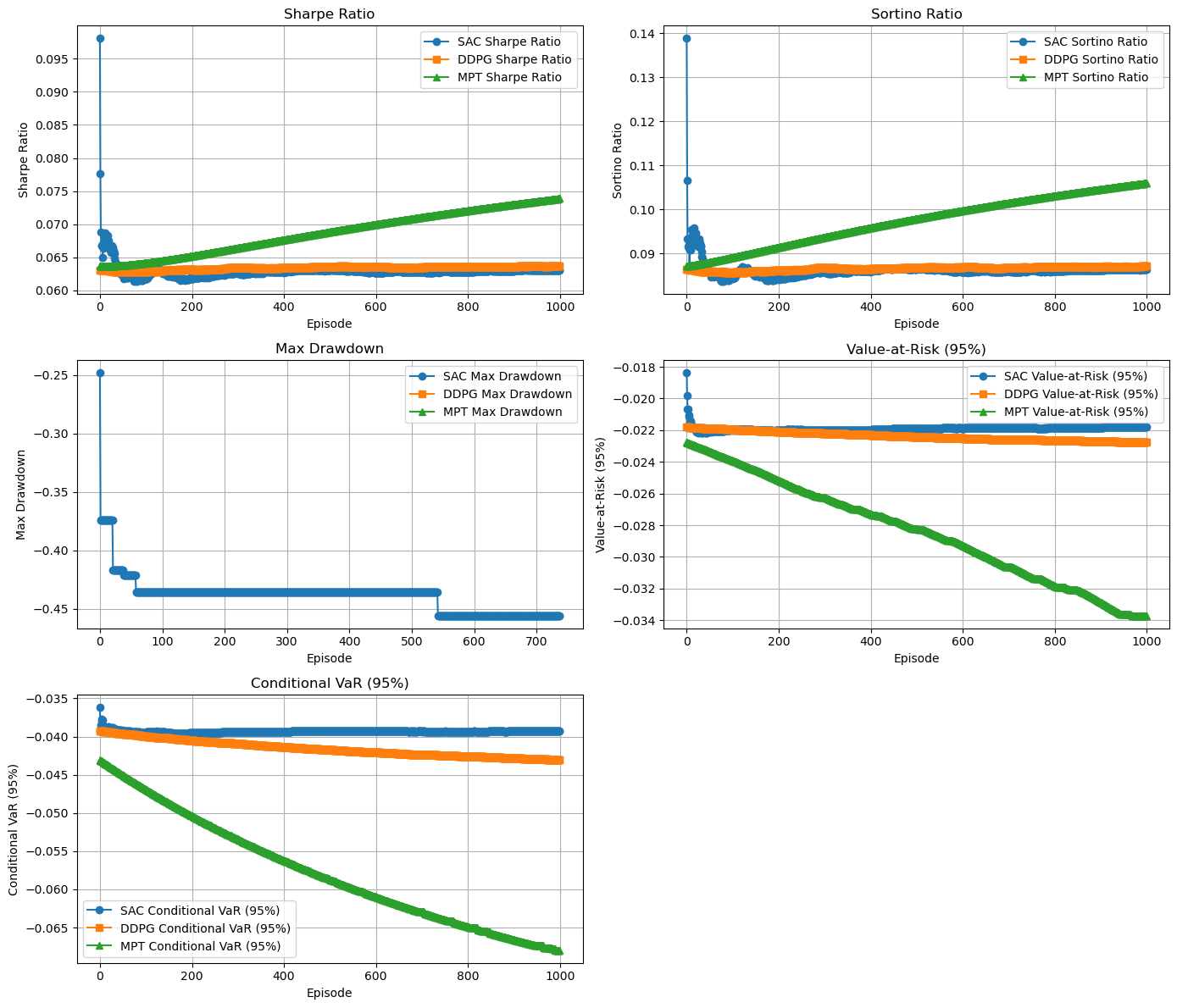}
\caption{Episodic comparison of Sharpe ratio, Sortino ratio, Maximum Drawdown, VaR (95\%), and CVaR (95\%) for SAC and DDPG algorithms.}
\label{Performance_Metrics_Comparison}
\end{figure}

Figure~\ref{Performance_Metrics_Comparison} illustrates episodic learning curves for Sharpe ratio, Sortino ratio, Maximum Drawdown, VaR, and CVaR. The DDPG agent improves gradually but exhibits higher tail risk, as shown by lower VaR and CVaR values. SAC, on the other hand, achieves smoother convergence with consistent downside risk control, demonstrating the stabilizing benefits of entropy regularization.

\begin{figure}[!htb]
\centering
\includegraphics[width=1.00\textwidth]{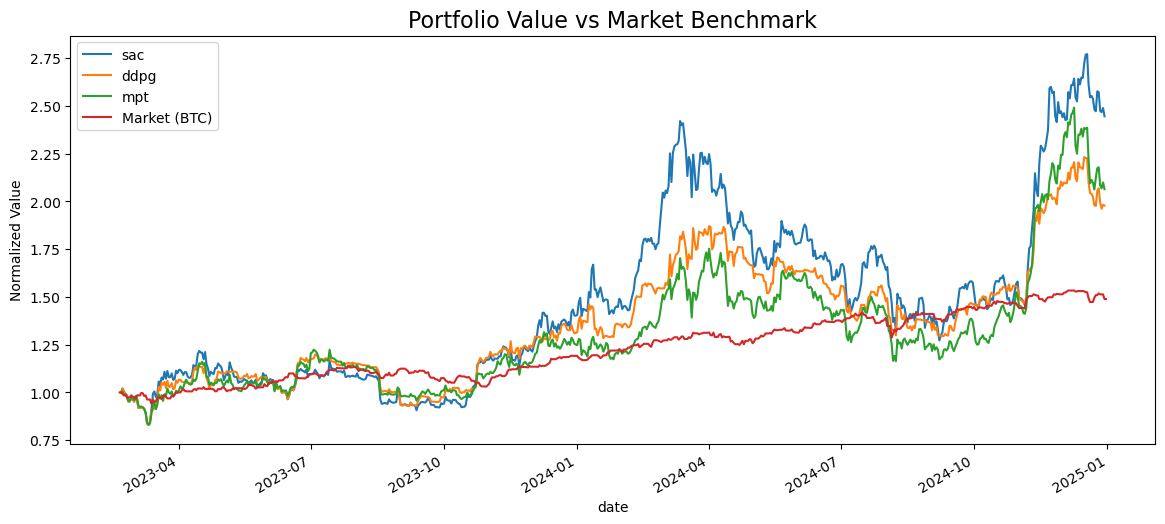}
\caption{Normalized portfolio values for SAC, DDPG, and MPT across BTC, ETH, LTC, and DOGE, compared with Bitcoin during the test period.}
\label{Portfolio_Value_Comparison}
\end{figure}

Figure~\ref{Portfolio_Value_Comparison} shows normalized portfolio values for SAC, DDPG, and the MPT strategy over the test period, benchmarked against Bitcoin. SAC demonstrates the strongest cumulative return, nearly tripling the initial investment. DDPG exhibits higher volatility and less consistent performance. The MPT strategy produces stable but slower growth. The results emphasize the adaptability of SAC and the effectiveness of reinforcement learning in dynamic market conditions.

Table~\ref{tab:test_results} summarizes the test-period results. SAC achieves the highest final portfolio value ($2.7627$), the largest mean log return, and the best risk-adjusted metrics (Sharpe = $0.0673$, Sortino = $0.1093$). It also exhibits lower maximum drawdown and reduced tail risk compared to DDPG and MPT.

DDPG shows negative mean returns and higher drawdowns, indicating instability in volatile environments. MPT produces conservative but steady results. SAC’s diversified asset allocation further supports its robustness in nonstationary markets.

\begin{table}[h!]
\centering
\caption{Performance comparison of DDPG, SAC, and MPT on the test dataset.}
\label{tab:test_results}
\renewcommand{\arraystretch}{1.25}
\setlength{\tabcolsep}{6pt}
\begin{tabular}{lccc}
\toprule
\textbf{Metric} & \textbf{DDPG} & \textbf{SAC} & \textbf{MPT} \\
\midrule
Final Portfolio Value & 1.969214 & 2.762704 & 2.035221  \\
Mean Log Return & -0.001463 & 0.001492 & 0.001043  \\
Standard Deviation & 0.022354 & 0.027567 & 0.023023 \\
Sharpe Ratio & 0.013126 & 0.067340 & 0.026794 \\
Sortino Ratio & 0.018515 & 0.109281 & 0.037859 \\
Maximum Drawdown & -0.681602 & -0.409029 & -0.719820 \\
VaR (95\%) & -0.040353 & -0.041248 & -0.039025 \\
CVaR (95\%) & -0.060043 & -0.057676 & -0.057710 \\
\midrule
Avg. Weight (BTC) & 0.515419 & 0.100008 & 0.202570 \\
Avg. Weight (ETH) & 0.010000 & 0.393539 & 0.233725 \\
Avg. Weight (LTC) & 0.048458 & 0.606461 & 0.224882 \\
Avg. Weight (DOGE) & 0.436123 & 0.101000 & 0.189328 \\
\bottomrule
\end{tabular}
\end{table}

\subsection{Conclusion and Future Work}

This study proposed a deep reinforcement learning framework for dynamic cryptocurrency portfolio management using SAC and DDPG, enhanced with LSTM networks for temporal feature extraction. A Markowitz MPT baseline was implemented for comparison. Historical data from 2016--2024 were used to train and test the models under realistic conditions.

Empirical evaluations showed that both RL-based agents outperform the classical MPT benchmark, with SAC demonstrating superior risk-adjusted performance, stability, and adaptability in volatile environments. DDPG achieved competitive returns but displayed greater sensitivity to market noise and hyperparameter initialization.

Challenges included overfitting in low-liquidity assets, data quality issues, and computational demands—particularly for SAC due to entropy-regularized updates. Despite these challenges, the results highlight the potential of reinforcement learning for robust and adaptive portfolio management in nonstationary markets.

Future research may incorporate multimodal data sources (e.g., sentiment, blockchain metrics), explore advanced architectures such as transformers, evaluate scalability to larger asset universes, and improve model interpretability using explainable AI techniques.

\section*{Disclosure Statement}
The author declares no conflicts of interest.

\section*{Funding}
This research received no external funding.



\end{document}